# Batch is back: CasJobs, serving multi-TB data on the Web

William O'Mullane,
Nolan Li,
María Nieto-Santisteban,
Alex Szalay,
Ani Thakar
*The Johns Hopkins University*

Jim Gray
*Microsoft Research*

# Batch is back: CasJobs, serving multi-TB data on the Web

William O'Mullane, Nolan Li, María Nieto-Santisteban, Alex Szalay, Ani Thakar
*The Johns Hopkins University*
*womullan@jhu.edu +1 410 516 4908*
Jim Gray
*Microsoft Research*

**Abstract:** The Sloan Digital Sky Survey (SDSS) science database describes over 140 million objects and is over 1.5 TB in size. The SDSS Catalog Archive Server (CAS) provides several levels of query interface to the SDSS data via the SkyServer website. Most queries execute in seconds or minutes. However, some queries can take hours or days, either because they require non-index scans of the largest tables, or because they request very large result sets, or because they represent very complex aggregations of the data. These "monster queries" not only take a long time, they also affect response times for everyone else – one or more of them can clog the entire system. To ameliorate this problem, we developed a multi-server multi-queue batch job submission and tracking system for the CAS called CasJobs. The transfer of very large result sets from queries over the network is another serious problem. Statistics suggested that much of this data transfer is unnecessary; users would prefer to store results locally in order to allow further joins and filtering. To allow local analysis, a system was developed that gives users their own personal databases (MyDB) at the server side. Users may transfer data to their MyDB, and then perform further analysis before extracting it to their own machine. MyDB tables also provide a convenient way to share results of queries with collaborators without downloading them. CasJobs is built using SOAP XML Web services and has been in operation since May 2004.

## 1 Sloan Digital Sky Survey - SkyServer

The SkyServer[1] has been available on the Internet since June 2001. It provides public access to the Sloan Digital Sky Survey (SDSS[2]) catalogs -- both the optical and spectroscopic data. The raw SDSS pixel data will approach 50 TB and is available from file servers at Fermilab. The catalogs are derived from this raw data. The current catalog database (DR3) of 230 million galaxies and stars and 530 thousand spectra covers 6,000 squared degrees. The catalogs are stored in a 1.6 TB SQL Server database. Subsequent data releases will double the database size. In addition to the catalog data, we have begun loading the points of all SDSS spectra in a separate database, adding another few hundred GB of data online. Hence the database will become much larger.

The SkyServer is an IIS Web server backed by a SQL Server database. It provides visual ways to explore the data and allows forms-oriented requests to query the database. Users can upload catalogs and cross-match them with the SDSS catalogs. Interactive SQL queries for more sophisticated data analysis are also supported.

### 1.1 SkyServer Statistics

The site now averages 2M hits per month; the traffic has been doubling every 15 months. Considering that it is running on two $10k servers, the site performs extremely well and most queries run quickly. However, in certain circumstances it has problems. Complex queries can swamp the system and erroneous queries may run for a long time but never complete.

On the public SkyServer, queries are limited to 10 minutes and answers are limited to 100,000 rows. On the professional astronomer and collaboration[3] sites, the limits are 1 hour of elapsed time and answers sets of 500,000 rows or less. Most queries run within seconds. Execution times and result set sizes follow a natural power law (see Figure 1). Hence there is no obvious point at which queries could be cut off. All SkyServer queries run at the same priority- there is no ranking or "nice" scheduling system built into SQL Server (or any other DB products). While this may not be a problem in itself, long queries can slow down the system, causing what should be quick queries to take much longer. Some queries return large answer sets to a user over an Internet connection.

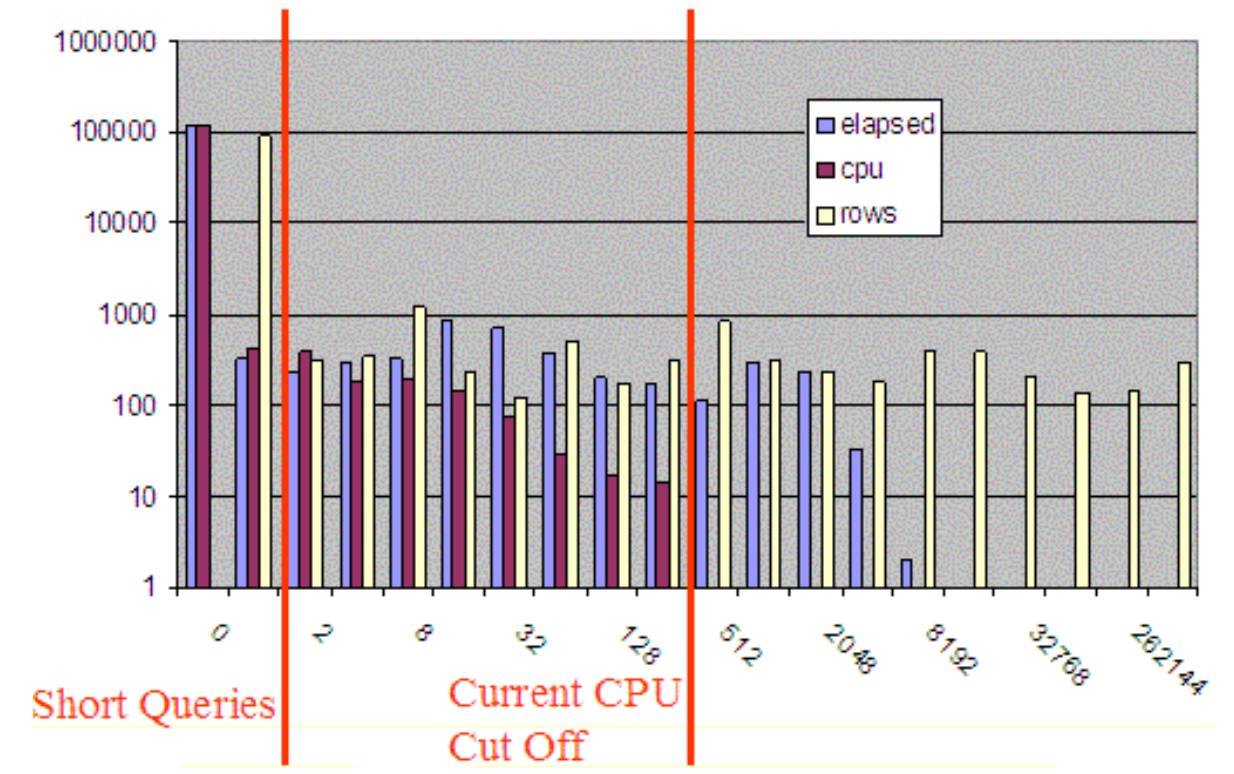

**Figure 1.** Log-log plot of Frequency of queries using n seconds, *n* cpu seconds, or returning n rows over 100 hour period. The curves suggest power laws – the products of size and frequency are nearly constant.

[1] http://skyserver.sdss.org
[2] http://www.sdss.org

Considerable time is spent transferring such data. Web server and database server resources are consumed during the transfer. We have seen as many as twelve million rows (20GB) downloaded in one hour. These large transfers are often unnecessary; the data are often intermediate results used only to make comparisons against a small local set, or used in the next analysis step.

Figure 2 shows the monthly CAS usage as logged on the SkyServer website, and compares the website hits with the number of SQL queries executed per month. We only started logging SQL queries in 2003. While the SkyServer traffic as a whole has been steadily increasing over the past 2 years, we have seen a significantly steeper increase in the SQL query usage since June 2003, indicating the increasing sophistication and SQL familiarity of SkyServer users. This was further corroborated by our SkyServer HelpDesk requests.

This rapidly increasing demand for direct SQL access to the database was also a motivator for seeking a system whereby users could run unlimited queries and have access to their own database workspace with the ability to test and refine complex queries iteratively.

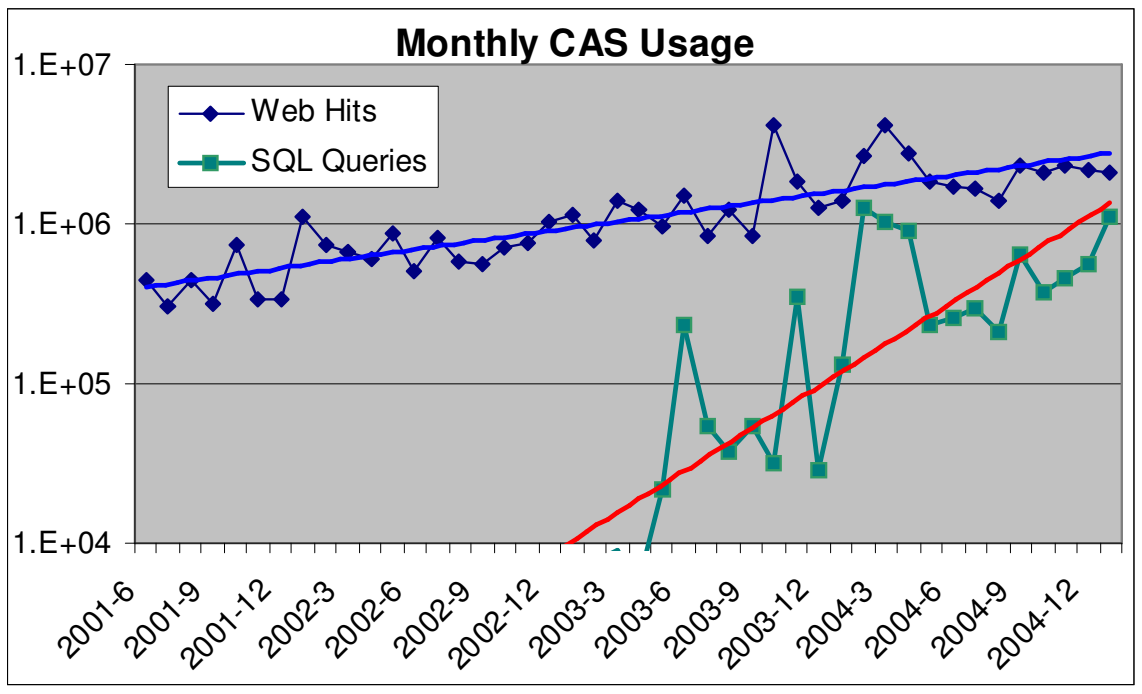

**Figure 2.** The monthly Web hits and SQL queries on the SkyServer website (portal to the CAS). SQL queries are becoming quite common.

# 2 Batch System

We developed the Catalog Archive Server Jobs System (CasJobs)[4] to address these problems. CasJobs allows multi-step analysis queries that pipeline intermediate results to one another, rudimentary load balancing across multiple machines, guarantees query completion/abortion, provides local table storage for users, and separates data extraction/download from querying. These features will be pertinent to the Virtual Observatory [6] as the SkyNode protocol matures [1][5].

## 2.1 Queues

CasJobs has multiple job queues based on expected query execution time. Jobs in the shortest queue are executed immediately; while jobs in all other queues are executed sequentially (limited concurrent execution is allowed). CasJobs currently runs with a one minute (shortest) queue and multiple five-hundred-minute queues.

CasJobs may have as many queues of different lengths as we wish. Jobs may be *auto-promoted* from one queue to the next if they exceed the queue's quantum but most users find auto-promotion confusing. Simply having one long and one short queue is more understandable, and has performed adequately for the moment.

Query execution time is limited by the time assigned to a particular queue. Several machines are dedicated to processing the workload. Each machine has a replica of the SDSS database and pulls work from the work queues. Hence there are no longer ghost jobs hanging around for days (because of the time limits) and long queries no longer hinder execution of shorter ones. A job may take only as long as its queue time limit, and different types of jobs are executed on different machines.

## 2.2 MyDB - Local Storage

Queries submitted to the longer queues must write results to a private local database, known as MyDB, using the standard *into* syntax e.g.

```
SELECT TOP 10 *
INTO MyDB.rgal
FROM galaxy
WHERE r < 22 AND r >21
```

The MyDB idea is similar to the AstroGrid MySpace [8] concept; however MySpace exists only for files at this time – MyDB is a full-blown SQL Server database with all the data definition, data manipulation, and programming abilities of SQL. The user can define stored procedures and programs in MyDB and use these procedures in analysis queries.

CasJobs creates a SQL Server database for the user dynamically the first time MyDB is used in a query. Upon creation, appropriate database links and grants are made such that the database will work in queries on any SDSS database. The name of the physical MyDB database is stored in a column of the Users table of the administrative database. The identifier of the machine where the database physically resides is stored in a separate column – this is a foreign key to the actual server details in the

[4] http://casjobs.sdss.org/casjobs

MyDBServers table. In this way individual MyDBs, or all of the MyDB's on a particular machine, may easily be moved to other machines MyDB. The MyDBServers table contains a limit to the number of MyDBs allowed on that machine. Multiple MyDB servers may be configured in this table and the MyDBs will be spread over these hosts.

Since the MyDB is a normal database the user may perform joins and queries on tables in MyDB as with tables in any other SDSS database. The user is responsible for this space and may drop tables from it to keep it clear. The MyDB screen (Figure 3) gives the user an overview of his MyDB. It shows him his tables and other resources. Users are initially given 500MB as the limit for their MyDB, but this is configurable on a system and per user basis. An administrative screen is available to administrators to change these settings.

## 2.3 Functions & Stored Procedures

Queries are a useful research tool for astronomy. However there is generally some further processing required to archive a final result. We would like to enable moving some of the processing closer to the data. The closest of course one may come is to have a stored procedure or function. We allow creation of stored procedures and functions in the users MyDB. Hence a user may issue a statement such as:

```
CREATE FUNCTION sq(@X BIGINT )
RETURNS BIGINT
BEGIN
RETURN @X*@X
END
```

Then the user can write queries like:

```
SELECT top 1000 dbo.sq(ra)
FROM dr2.PhotoObjAll
```

More generally, the user can write any Transact-SQL program and either runs it or installs it as a function or stored procedure in MyDB. On the MyDB screen (Figure 3) the user may see which tables, functions and procedures are available in the MyDB or any other database.

## 2.4 Privileges

CasJobs implements a simple role-based privilege system. Each user has a list of roles associated with their account. The roles are arbitrary strings. Internally we use meaningful strings such as "ADMIN", "QUERY", "MyDB" and "GROUP" for some of the system privileges. Each queue may also have a privilege associated with it, only users with the same privilege in their role string will then see that queue. Hence a certain database may be hidden by assigning privileges to the queues associated with that database. This is important, as some of the latest data is only available to the collaboration members for a period of time. Assigning "collab" privilege to the queues for the new data releases means that public users do not see this restricted data before they should.

## 2.5 Security and Groups

Today each user has a password. We are exploring other authentication mechanisms (e.g. certificates and Web services standards.)

A user may wish to share data in his MyDB. Any user with appropriate privileges may create a group and invite other users to the group with read-only or read-write privileges. An invited user may accept being part of the group. A user may then publish any of his MyDB tables to the groups of which he is a member. Other group members may access these tables by using a pseudo database name consisting of the word *group* followed by the *userid* of the other user followed by the table name e.g. if the *Hopkins* user published the table *rgal* and you were in a group with *Hopkins* you may access this table using *GROUP.Hopkins.rgal*.

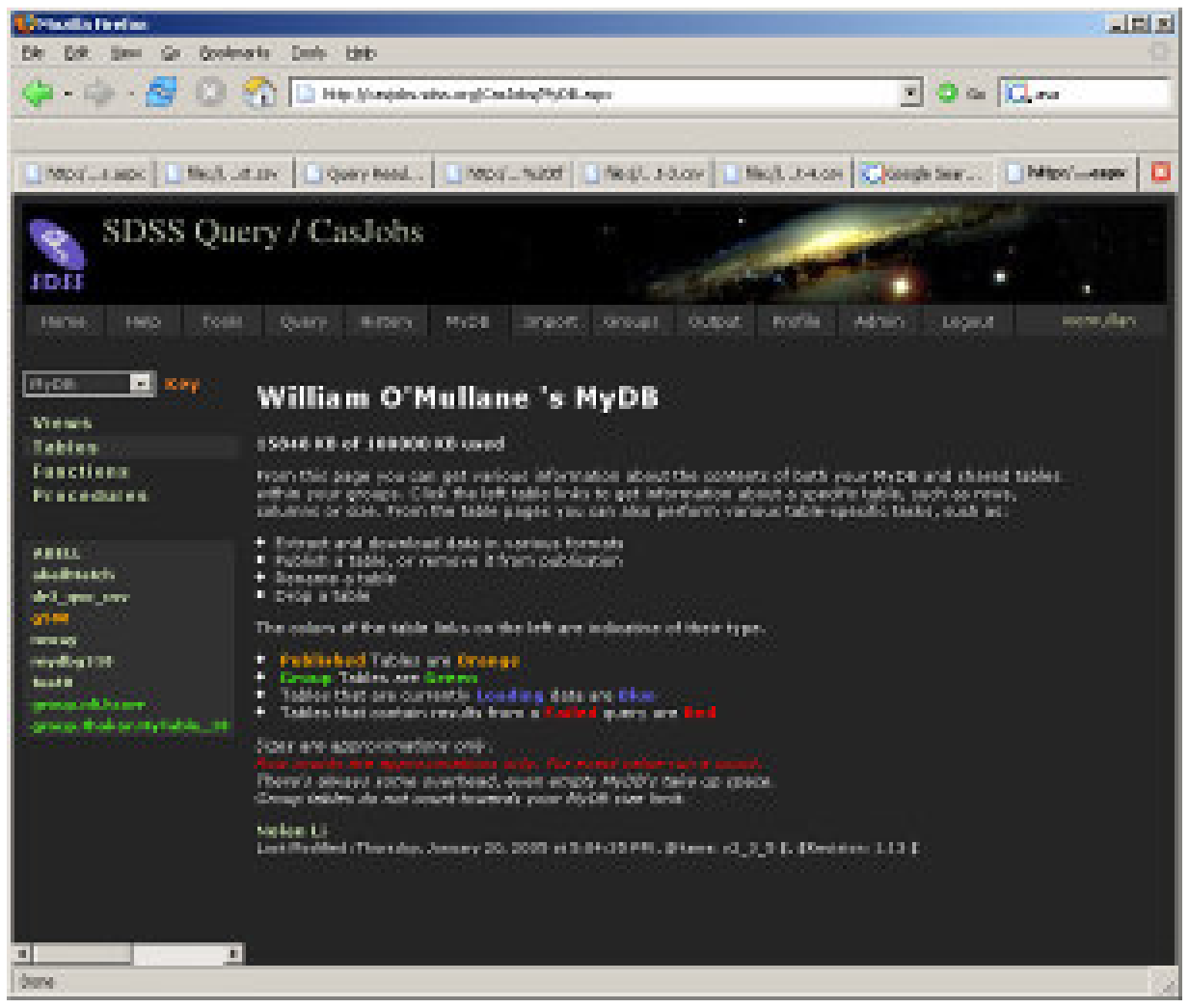

**Figure 3.** MyDB screen gives the user an overview of his database and gives shortcuts to common administrative and analysis tasks.

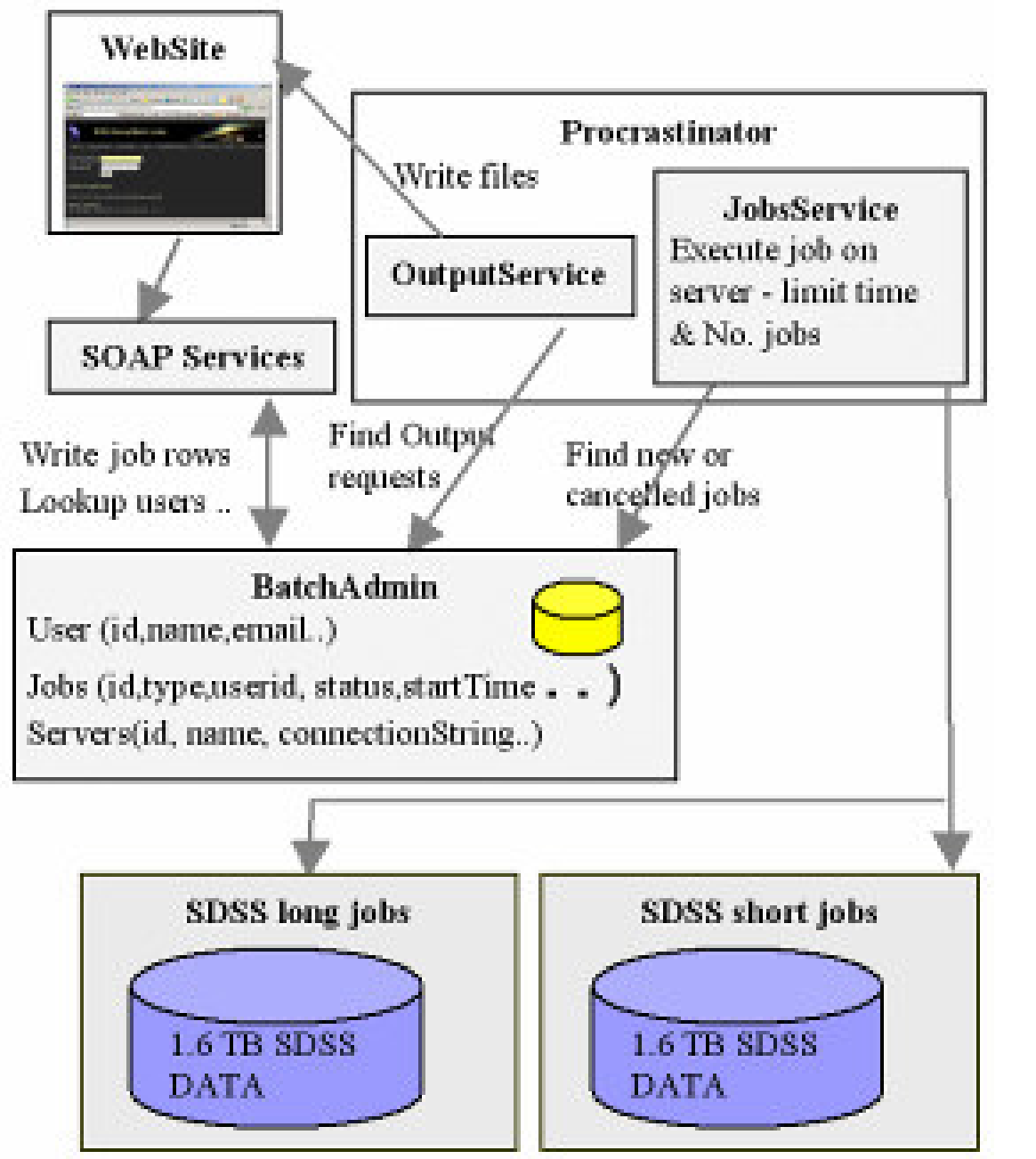

**Figure 4.** The CasJobs system architecture has a web front-end that issues SOAP calls to create jobs and query their status. The procrastinator schedules the jobs against a pool of servers – some dedicated to short jobs and some to long jobs. Results are deposited in MyDB or in files that can be downloaded via the Internet.

## 2.6 Import/Export Tables

Tables from MyDB may be requested in FITS [2], CSV, or VOTable[5] format. Extraction requests are queued as a special job type and have their own resources. Once the file extraction is done, a URL to the file location is put in the job record.

A user may also upload a data file as CSV or VOTable to their MyDB. A cross-match procedure (spGetNeighbors) available in SDSS databases has been made easily accessible on the MyDB screen. This procedure lets users do position-based neighbor comparisons between objects in the MyDB table and other SDSS tables. The ability to upload data and the group system have reduced the huge downloads from the server. This however requires a change in many astronomers thinking and work-habits.

[5] http://www.us-vo.org/VOTable/

## 2.7 Jobs

Apart from the short jobs, everything in the CasJobs system is asynchronous and requires job tracking. Each job has an entry in the *jobs* table of the administrative database. Submission of a job simply creates this entry. A windows service (The *Procrastinator*) runs a thread which wakes up periodically and scans the job table for new jobs for each target specified in the Servers table. If a server is not running its allowed number of jobs, then a thread is spun off for the job, the thread includes a timer which cancels the query and closes the connection if the job runs longer than the specified queue length. The entry in the jobs table is updated to show the job has started and will be updated once more to show whether it completed or failed. At this point the service will also mail the user concerning the job if they have selected that option in their profile. This thread also checks for jobs which have been marked "canceled" in the *jobs* table; such jobs are stopped.

The Procrastinator also runs a separate thread for File Export (output) jobs. This also wakes up periodically to scan the jobs table for output jobs. It creates the output files (sequentially) and updates the job entries. This process also scans the HTTP directory where files are written and removes files older than a configurable time (currently one week). Figure 4 gives an overview of the main components involved. Currently, the Web site and Procrastinator run on one machine, the BatchAdmin is on a machine with the MyDBs and the SDSS databases are on dedicated hardware. The architecture is flexible allowing for the components to easily be moved to various machines.

The job system has a rudimentary password-based authorization system for users and groups. It has an interface that allows Web applications to submit jobs, manage their MyDB (see Figure 3), query a job's status, and cancel it. A user may list all previous jobs and get the details of status, time submitted, started, finished etc. The user may also resubmit a job. The job history page is shown in Figure 5.

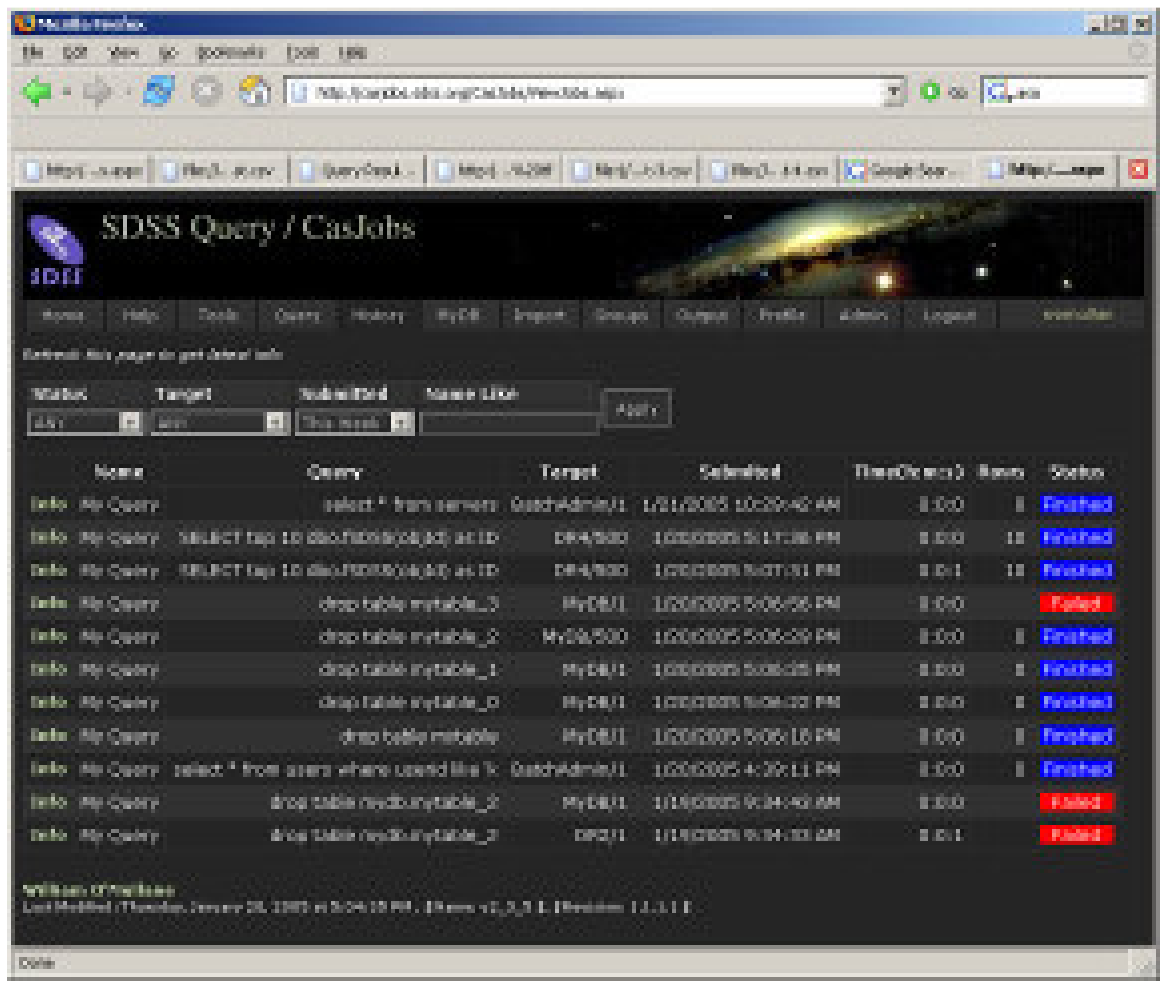

**Figure 5.** The job history screen shows past queries and their outcomes.

## 2.8 Rewriting the Query

To support the MyDB and GROUP table prefixes, queries are rewritten before execution to translate these aliases into internal system names. MyDB is contextual based on the user issuing the command. It is replaced with the real name of that user's database. We discovered that the *into* syntax is expensive when linked databases are involved. It appears that the entire query is executed, the result put in memory on the server, then moved to the target machine (put in memory), and finally written to the target database. To circumvent this, the "into" clause is actually removed from the query, the query is executed and a SQLDataReader is created, resulting effectively in a database cursor. On the target end, a table is created in MyDB and an insert statement is prepared. Each row is then read from the source and inserted into the target. For any moderate size query this is as efficient as, or more efficient than, using SQLServer's *into* syntax. It has the added advantage of never using much memory and allowing the user to see initial results of the query in MyDB as they arrive.

For the GROUP prefix we know the table is read-only. This is simply a matter of looking up the correct database and table and replacing it in the query. Permissions are also checked at this point, i.e., if the table was not published to the group then the substitution is not done.

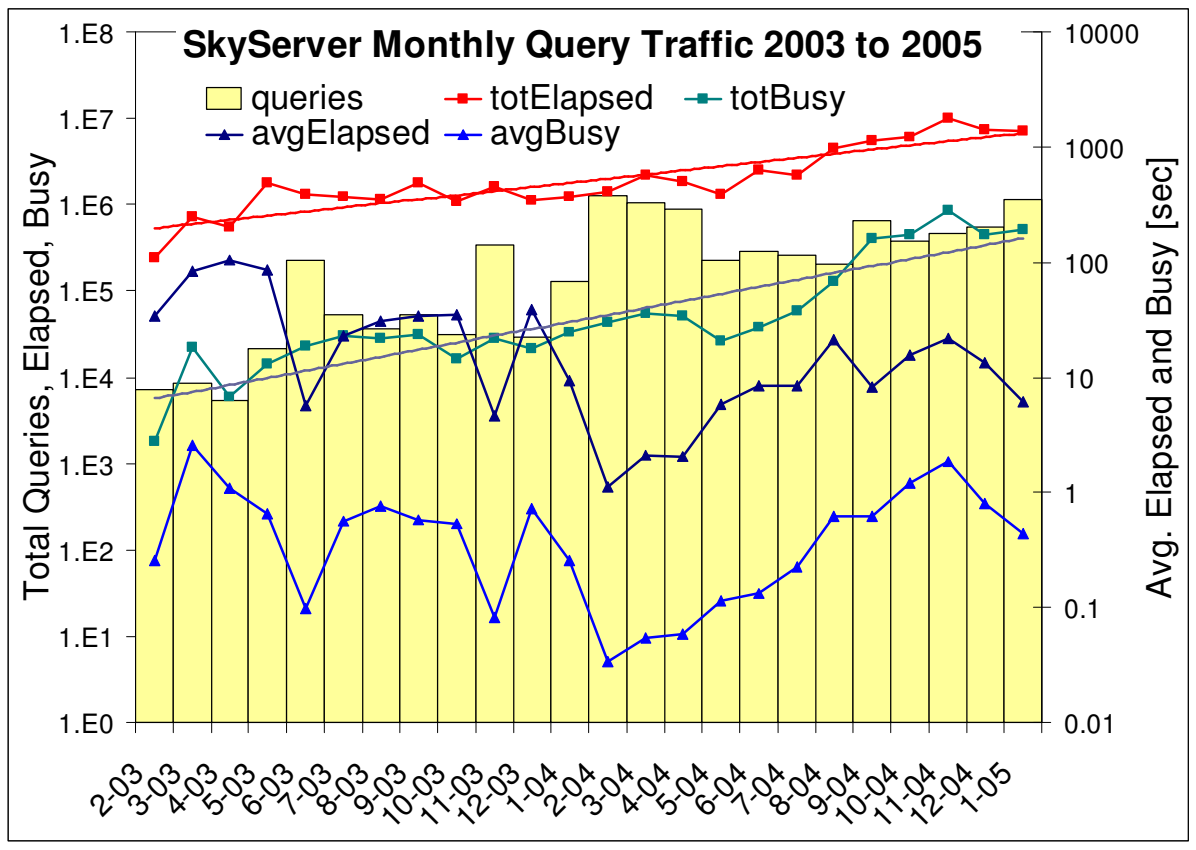

**Figure 6.** SkyServer monthly query traffic has grown over the last 2 years. Elapsed time has improved due to better hardware and due to offloaded CasJobs even though the database size has increased 10x and the workload has increased 100x.

The administrative database has a table of regular expressions. For example we check to see if users are trying to execute system stored procedures or drop tables outside MyDB. These expressions are run over the query and may cause it to be rejected. This allows us to check for suspicious activity in queries.

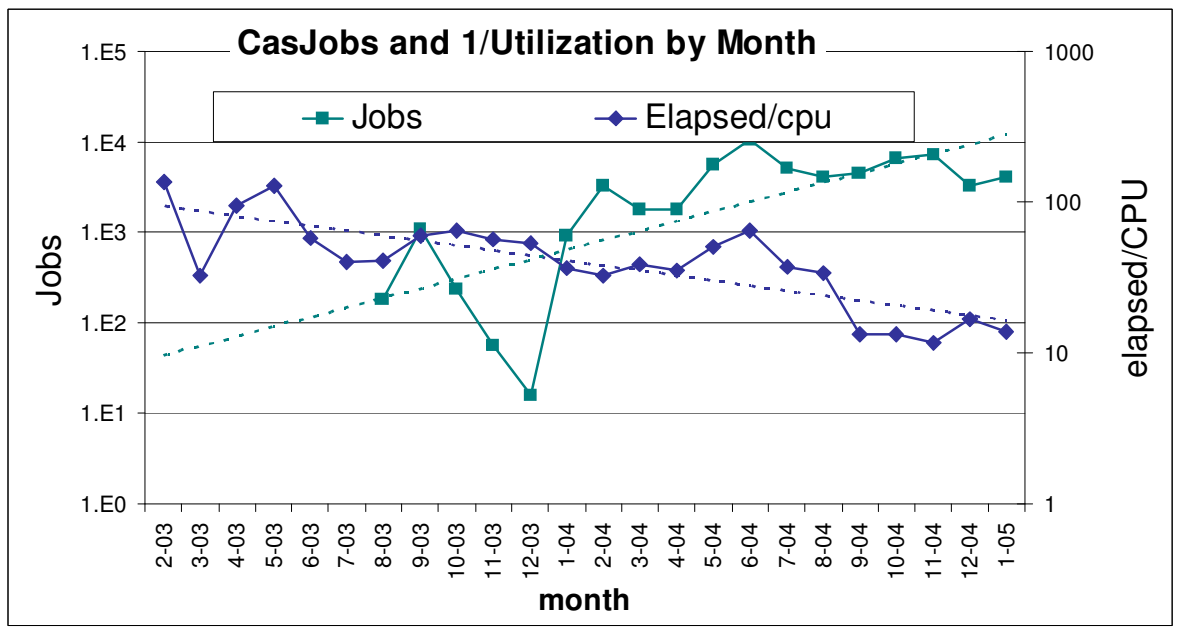

**Figure 7.** CasJobs has increased by 100x in the last year to about 5,000 jobs per month. In that time, CPU utilization has gone from 1% to 10% -- the queries are still IO bound, but much less so now that answers can be fed to MyDB.

## 2.9 Analyzing the Impact of CasJobs

CasJobs has been in service for over a year and by the end of January 2005 had run over sixty thousand jobs. We can analyze its impact on the data access patterns by comparing data access patterns and system loads before and after the introduction of CasJobs. Figure 6 shows the monthly average and total elapsed and CPU busy times for SkyServer queries. In that time, the SkyServer ran over 8.5 million database queries. The database grew from 100GB to 1TB – so each query had to process a 10x larger database at the end of the period. The plot shows the total number of queries per month for reference. Trend lines are shown for the total elapsed and busy times. The number of queries per month is about an order of magnitude higher since the introduction of CasJobs (Fall 2003). The system is IO-bound by a large factor (the vertical distance between the elapsed and busy CPU times), but this factor is decreasing: the trend lines for the total elapsed and busy times are converging with time.

This relationship is clearer in Figure 7, which shows the CPU utilization (actually it plots the reciprocal $1/\mu$) and compares it with the number of CasJobs queries per month. The utilization has improved from 1% to 10% while CasJobs usage has increased to several thousand jobs per month. CasJobs and MyDB have given much better CPU utilization and the system is handling many more queries with the CasJobs/MyDB service in place.

Figure 8 shows the work generated by the various users. Most of the users ran some huge jobs. The number of jobs per user follows a power law, and the total consumption also seems to follow a power law. Users are just becoming familiar with the use of CasJobs and MyDB. It will be interesting to track this growth over the next few years. The users with over a thousand jobs have discovered how to submit jobs programmatically.

### 2.10 Ferris Wheel

A future experiment will batch full table scan queries together. Theoretically we may piggyback queries in SQL Server so that a single sequential scan is made of the data instead of several. Ideally we would like to not have to wait for a set of queries to finish scanning to join the batch. Rather we would like some set of predefined entry points where a new query could be added to the scan. Conceptually one may think of this as a Ferris wheel where no matter which bucket you enter you will be given one entire revolution.

## 3 SOAP Services

We have found that SOAP services provide a clean API for building modular distributed systems. In contrast to some of the existing myths about the performance of SOAP, our experience has shown the SOAP overhead to be tolerable and the performance to be adequate with multi-GB datasets. The CasJobs website is based on a set of SOAP services. Any user may access these services directly using a SOAP toolkit in their preferred programming language. At JHU we have used Python, Java (AXIS) and C# clients for Web services successfully. Others have written Perl clients. More information on this is available at the International Virtual Observatory Alliance (IVOA) Web site[6].

A command line package for accessing CasJobs was developed in Java[7]. This was done as an example of how to programmatically access the services provided by CasJobs as well as being a useful tool and demonstrating interoperation between .NET and Apache AXIS.

### 3.1 Web Services: Stateful or not?

The question of how much state is in a Web service is a much-discussed issue at the moment. The Web services Grid Application Forum has it as a core issue [5]. Technically the CasJobs Web services are stateless. We do not use any Web service or Grid technology to track state in the SOAP call. Each SOAP call could be answered by any server hosting the CasJobs software and able to access the CasJobs database. Obviously this is not the whole picture. CasJobs has a database of jobs belonging to a user and manages privileges and authorization on an individual basis. We choose however to follow the template of most e-Commerce sites, and implement that state within our system using database technology, much as e-Commerce shopping carts are generally implemented. In each call the WSID of the user is passed as a parameter, internally we use that to set the context of the message and use to find the state in the SQL database. Hence the service is stateful but without any library overhead.

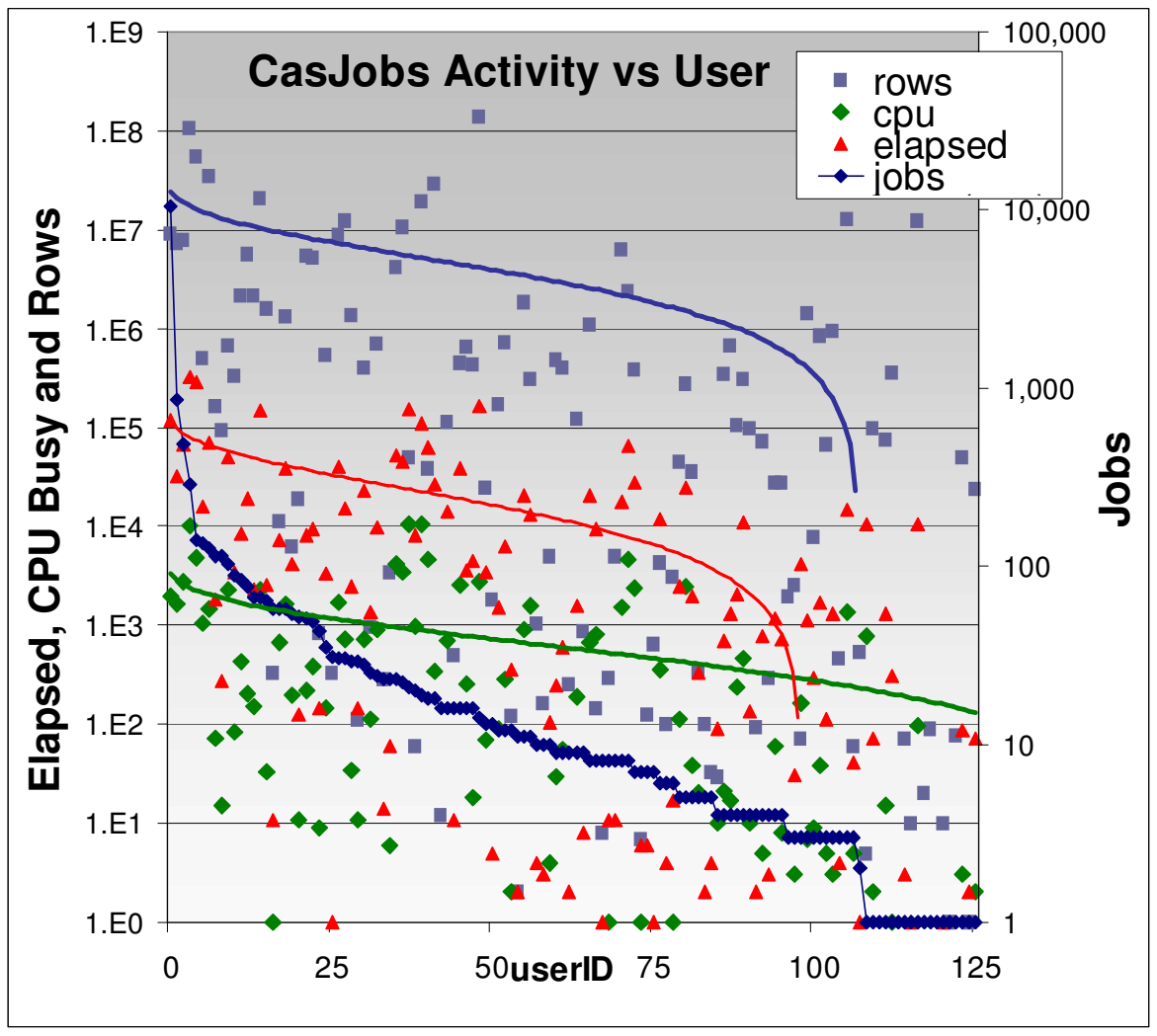

**Figure 8.** The 125 CasJobs users follow a power law in the number of jobs, and also the number of rows, cpu, and elapsed time.

One of us (WOM) has worked extensively with Enterprise Java Beans (EJB) where a similar debate was rife, personal experience showed that stateless EJBs backed by a similar system to the above proved far more efficient than their stateful counterparts.

### 3.2 Security

CasJobs is not security conscious. Typically astronomical data is freely available; CasJobs uses simple password protection on the Web site more to identify users than to provide a high level of security. WS-Security [3] would allow us to seamlessly put a certificate layer on this. Initial attempts at interoperation using WS-Security between .NET and Java AXIS were disappointing. WS-Security is now working correctly. It is not clear however that we want to or need to burden our users with getting and using certificates at this time.

[6] http://www.ivoa.net/twiki/bin/view/IVOA/WebgridTutorial
[7] http://casjobs.sdss.org/CasJobs/casjobscl.aspx

# 4 Portability

The queues in CasJobs are in fact simple database connections. To add another database to the interface requires adding a row to the Servers table. The only other requirement is to create a database link between any machine with MyDB databases and the machine with the database. Apart from this small effort the system is not tied in anyway to SDSS. We successfully put the system on a server at another institute and had it serving up data from one of their databases in about one hour.

# 5 Conclusion

CasJobs provides a mechanism for sharing very large databases through Web services. It tackles the problem of large queries swamping the system as well as run away queries that never finish. The Web portal and Web service interfaces provide users with a sophisticated front end for managing their queries as well as their personal MyDB. This has allowed scientists to interact with and analyze the large volume of SDSS data without copying it to their local institutes.

# 6 Bibliography

[1] Budavári, T., et al. "Open SkyQuery – VO Compliant Dynamic Federation of Astronomical Archives" in [4]

[2] Hanisch R. J., et. al., "*Definition of the flexible image transport system (FITS).*" Astronomy and Astrophysics, 376 pp. 359--380, September 2001.

[3] *OASIS Web Services Security*: http://www.oasis-open.org/committees/wss.

[4] Ochsenbein F., Allen M. and Egret D. *Astronomical Data Analysis Software and Systems* XIII ASP Conference Series Vol. 314, 2004.

[5] Parastatidis. S, Webber J., Watson P, Thomas Rischbeck, "A Grid Application Framework based on Web Services Specifications and Practices," Journal of Concurrency and Computation: Practice and Experience. To appear 2005.

[6] Szalay A. S., "The National Virtual Observatory", Astronomical Data Analysis Software and Systems X, ASP Conference Proceedings, Vol. 238, 2001

[7] Yasuda, N., et al., "Astronomical Data Query Language: Simple Query Protocol for the Virtual Observatory" in [4]

[8] Walton, A., et al "AstroGrid: Initial Deployment of the UK's Virtual Observatory" in [4]